\documentclass{article}
\usepackage{spconf,amsmath,graphicx}

\usepackage{amsmath,amsfonts,amssymb}
\usepackage{diagbox}
\usepackage{adjustbox}
\usepackage{xcolor}
\usepackage{graphicx}
\usepackage{enumitem}
\usepackage{algorithm}
\usepackage{algpseudocode}

\setlist{nosep, leftmargin=14pt}

\usepackage{mwe} 

\title{Detecting and measuring human gastric peristalsis using magnetically controlled capsule endoscope}
\name{Xueshen Li$^{1,2}$, Yu Gan$^{1}$, David Duan$^{2}$, and Xiao Yang$^{2}$}
\address{$^{1}$Department of Biomedical Engineering, Stevens Insitute of Technology, Hoboken, USA\\
$^{2}$AnX Robotica, Plano, USA}

\begin{document}
%
\maketitle
%
\begin{abstract}
Magnetically controlled capsule endoscope (MCCE) is an emerging tool for the diagnosis of gastric diseases with the advantages of comfort, safety, and no anesthesia. In this paper, we develop algorithms to detect and measure human gastric peristalsis (contraction wave) using video sequences acquired by MCCE. We develop a spatial-temporal deep learning algorithm to detect gastric contraction waves and measure human gastric peristalsis periods. The quality of MCCE video sequences is prone to camera motion. We design a camera motion detector (CMD) to process the MCCE video sequences, mitigating the camera movement during MCCE examination. To the best of our knowledge, we are the first to propose computer vision-based solutions to detect and measure human gastric peristalsis. Our methods have great potential in assisting the diagnosis of gastric diseases by evaluating gastric motility.
\end{abstract}
\section{Introduction}
Gastric motility is a process by which food travels through the digestive tract via a series of muscular contractions called peristalsis. Storage of ingesta, mixing and dispersion of food particles, and expulsion of gastric contents into the duodenum reply on gastric motility and activity \cite{Jean.1999}. Traditional methods of evaluating gastric motility, such as manometry, gastric emptying scintigraphy, and electrogastrography have their limitations. 
Manometry involves intranasal intubation protocols, which may cause discomfort to patients and lead to use of sedation \cite{Christian.2018}.
Nuclear medicine is required for gastric emptying scintigraphy, which leads to the risk of radiation exposure to patients \cite{PalashKar.2015}. 
Electrogastrography has many variations in the recording system. The lack of standards in analysis methods makes the electrogastrography hard to interpret \cite{CHANG.2005}.
Magnetically controlled capsule endoscope (MCCE) is an emerging tool for diagnosis of gastric diseases \cite{Carpi.2006}, possessing advantages of comfort, safety, and no anesthesia \cite{Zhang.2021b}. Moreover, MCCE provides real-time, true-color visualization of gastric environments that is easy to interpret. 
\begin{figure}[t]
\centering
\includegraphics[width=\columnwidth]{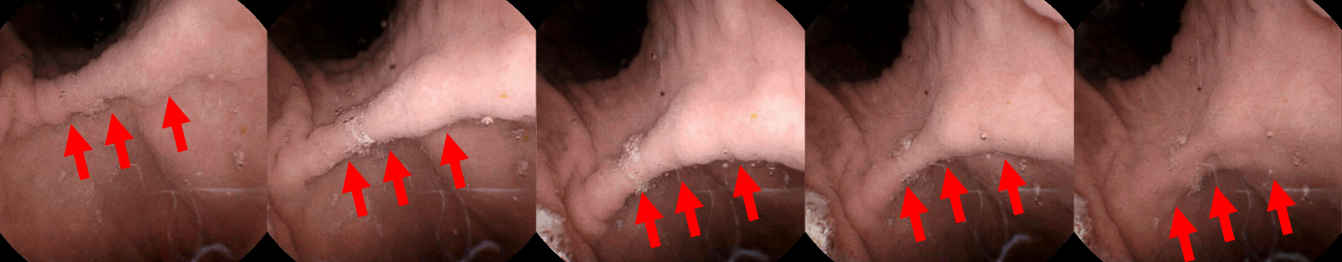}
\caption{A video frame sequence of gastric contraction wave captured by MCCE. The contraction wave has spatial features within single frames and temporal features across the frame sequence. 
The sequence is temporally undersampled by eight for better visualization. The red arrows highlight the contraction wave.}
\label{fig:waves}
\end{figure}

The median gastric examination and transit time using MCCE takes around one hour in clinical applications \cite{Liu.2022}. An example of gastric contraction wave captured by MCCE is shown in Figure \ref{fig:waves}. For evaluating gastric motility, consistent attention from doctors and experts are required during the whole examination process. 
There is a need to develop computer vision-based algorithms for analyzing MCCE video sequences and evaluating gastric motility.
During the gastric examination, the MCCE capsules have five controlled degrees of freedom (two rotational and three translational) and one uncontrolled degree of freedom (rotation along the long axis of the capsule).
Action recognition from moving cameras poses significant challenges \cite{Wu.2011}. To mitigate the sudden movement of the MCCE capsule camera in the six degrees of freedom, we develop a camera motion detector (CMD) for processing MCCE video sequences. We evaluate gastric motility from two aspects. First, we detect the presence of contraction waves in MCCE video sequences. Then, we measure the period of contraction waves. We use convolutional neural network (CNN) and long short-term memory (LSTM) for contraction wave detection. Moreover, we develop a periodical detector that measures the periods of human gastric contraction waves.

Using the CMD, we process more than 100,000 MCCE video frames. We generate a dataset with a stable camera view for training the CNN+LSTM model. We apply our algorithms to 30 MCCE video sequences, independent of training, for detecting and measuring the period of gastric contraction waves.  
Our methods have great potential in assisting the diagnosis of gastric diseases by evaluating gastric motility.
\section{METHOD}
\subsection{MCCE data collection}
MCCE video sequences were provided by the research department of Ankon Technologies Co Ltd (Shanghai, China). 
NaviCam\textsuperscript{\textregistered} MCCE system was used to collect video sequences from healthy internal volunteers in Ankon. 
The MCCE system consists of a swallowable capsule endoscope (11.8×27 mm), a guidance magnetic robot, a data recorder, and a computer workstation with softwares. 
The video frames were captured and recorded at two frames per second (fps).
The size of video frames were 480$\times$480 pixels.
\subsection{Camera motion detector design}
\begin{algorithm}
\small 
\caption{The camera motion detector}\label{alg:cap1}
 \hspace*{\algorithmicindent} \textbf{Input: } \textbf{Frame N}, \textbf{Frame N - 1}, and threshold \textbf{T}. \\
 \hspace*{\algorithmicindent} \textbf{Output: } A boolean variable \textbf{ifStable} for \textbf{Frame N}.
\begin{algorithmic}[1]
\State Calculate the residual image \textbf{R} between \textbf{Frame N} and \textbf{Frame N - 1}. 
\State Calculate histogram \textbf{H} of the residual image \textbf{R}.
\State Calculate \textbf{H}$_M$ by applying a mask \textbf{M} to the histogram.
\State Calculate score \textbf{S} by summing all the components in the masked histogram \textbf{H}$_M$.
\If{\textbf{S} $>$ \textbf{T}}
\State \textbf{ifStable} $=$ False
\Else
\State \textbf{ifStable} $=$ True
\EndIf 
\end{algorithmic}
\end{algorithm}


The CMD is capable of detecting whether or not a frame N is captured in a stable MCCE camera view. 
The details of our CMD are described in Algorithm \ref{alg:cap1}. 
A normalized Gaussian function with $\mu$ at 128 and $\sigma$ at 20 is adopted as the mask $M$. T is set to be 200. 
\subsection{Detecting human contraction waves using CNN, LSTM, and CMD}
\begin{figure}[t]
\centering
\includegraphics[width=\columnwidth]{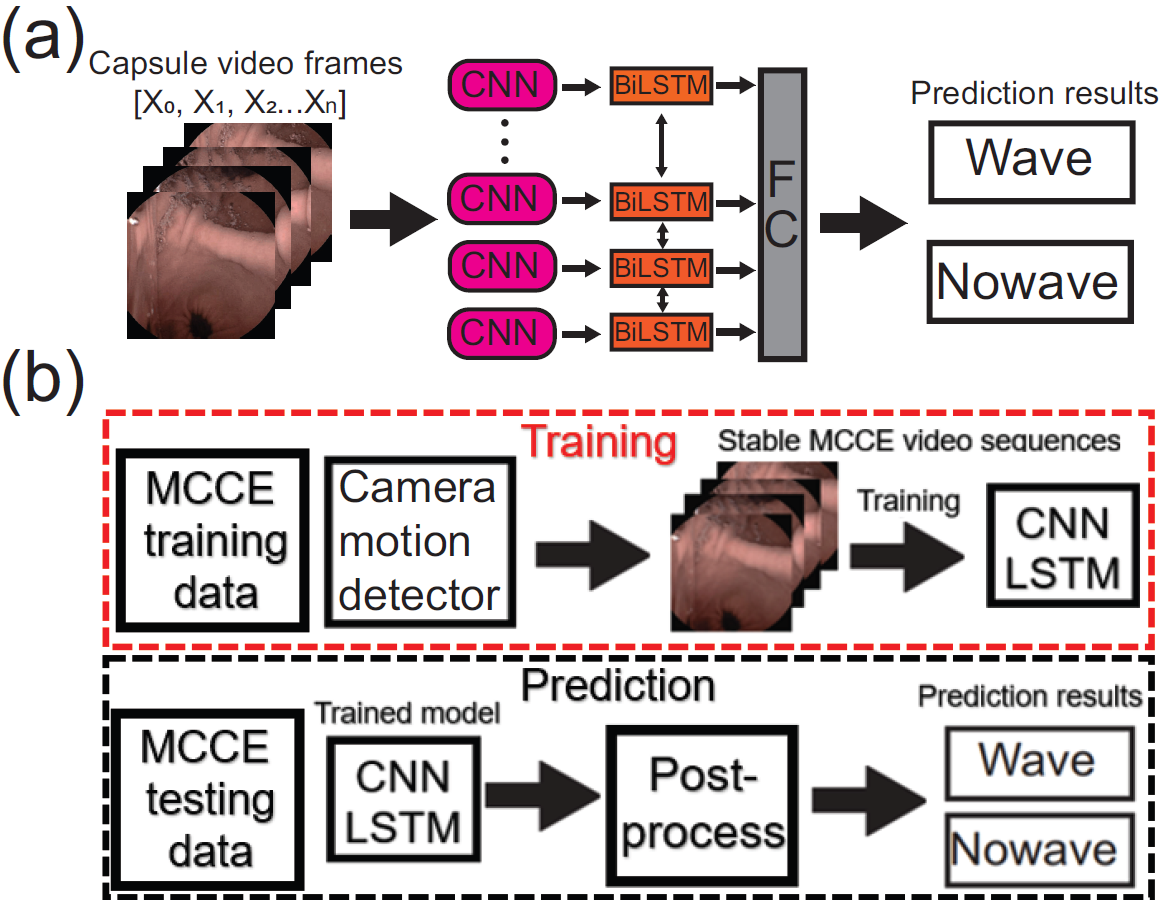}
\caption{(a) Design of our CNN+LSTM model for gastric contraction wave detection. N frames of capsule video records are fed to the CNN model. Then the extracted features from CNN are fed to LSTM cells. Outputs of LSTM cells are sent to a fully connected (FC) layer for binary classification of 'Wave' and 'Nowave'. We use EfficientNet-B7 \cite{Tan.2019} 
as the CNN. We implement bidirectional LSTM (BiLSTM) layers which acquire sequence information in both future and past directions. (b) Workflow of our ensemble of EfficientNet-B7, BiLSTM, and CMD for training and prediction stage of contraction wave detection.}
\label{fig:LSTMCNN}
\end{figure}
The human gastric contraction waves present features in both spatial and temporal domains. In the spatial domain, the waves have morphological shapes; in the temporal domain, the shape of waves changes over time. 
We use CNN architecture to extract time-domain features and LSTM architecture to aggregate the features in the temporal domain.
Moreover, we adopt bi-directional LSTM so the model is capable of capturing the temporal features of contraction waves in both future and past directions. The details of our implementation of CNN+LSTM model are shown in Figure \ref{fig:LSTMCNN} (a).

The workflow of training and prediction (testing) of the deep learning model is shown in Figure \ref{fig:LSTMCNN} (b). During training, the CMD serves as a pre-processing step to standardize input images as stable MCCE video sequences. The MCCE video sequences with unstable camera views will be removed. During testing, a similar idea of motion detection is involved in the post-processing step to select reliable prediction results of the CNN(EfficientNet-B7)+(Bi)LSTM model. If Frame N fails to pass the stable status detection (i.e., ifStable=False), the prediction results of the previous Frame N-1 will be adopted. This post-processing step is repeated until a valid frame that passes the stable status detection is found. As the starting location of imaging is selected by the operator of MCCE, we assume that the first frame is always stable. For post-processing, the CMD runs after the second frame as the first frame is always stable.
\subsection{Periodical detector for human contraction waves}
\begin{algorithm}
\small 
\caption{The periodical detector}\label{alg:cap2}
 \hspace*{\algorithmicindent} \textbf{Input: } \textbf{Intervals [1,2...I]}, \textbf{MCCE Sequence [1,2...N]},  threshold \textbf{$T_{l}$}, and \textbf{$T_{r}$}.\\
 \hspace*{\algorithmicindent} \textbf{Output: } The period \textbf{P} for human contraction wave captured by \textbf{MCCE Frames}.
\begin{algorithmic}[1]
\State Starting from MCCE \textbf{Frame 1}, use interval \textbf{i} to calculate \textbf{CMD} score $\textbf{S}_1$: $\textbf{S}_1$ = \textbf{CMD}(\textbf{Frame 1}, \textbf{Frame 1 + 1 $\times$ i}).
\State Calculating $\textbf{S}_2$:  $\textbf{S}_2$ = \textbf{CMD}(\textbf{Frame 1 + 1 x i}, \textbf{Frame 1 + 2 $\times$ i}).
\State Calculating $\textbf{S}_n$ until 1 + n $\times$ i $>=$ \textbf{N}.
\State Calculating the mean value $\textbf{S}^{mean}_i$ of [$\textbf{S}_1$, $\textbf{S}_2$...$\textbf{S}_n$] for interval $\textbf{i}$.
\State Repeating the previous steps, calculating $\textbf{S}^{mean}_i$ for \textbf{Intervals [1,2...I]}.
\State Finding the local minimal \textbf{P} of $\textbf{S}^{mean}_i$ between interval \textbf{$T_{l}$} and \textbf{$T_{r}$}.
\end{algorithmic}
\end{algorithm}
The human contraction waves present a periodical pattern \cite{Sarna.2013b}.
During clinical examination, the MCCE device may capture multiple contraction waves in a single frame. Thus, the wave periods cannot be simply measured by the presence of gastric contraction waves. We design a periodical detector for measuring periods of human contraction waves using MCCE video sequences. The details of the periodical detector are shown in Algorithm \ref{alg:cap2}. We set intervals from 5$s$ to 50$s$, with an incremental of 0.5$s$ (2fps). The $T_l$ is set to 10$s$ and $T_r$ is set to 40$s$. 
\section{EXPERIMENTAL DESIGN AND RESULTS}
\subsection{Training details}
We implement the EfficientNet-B7+BiLSTM model on PyTorch. The BiLSTM sequence is set to 5,10 and 20 frames. We use Adam optimizer for training for 100 epochs. The cross-entropy loss is adopted during training. The EfficientNet-B7 parameters are pre-trained on ImageNet. The first 5 epochs are warm-up epochs, in which the BiLSTM cells are frozen and only the EfficientNet-B7 model is trained. The learning rate is initialized as $10^{-4}$, followed by half decay for every 10 epochs. 
Each batch contains 8 MCCE video sequences.
The experiments are carried out in a single RTX 3080 GPU. The training procedure takes for approximately 4 hours.
\subsection{Evaluation setup and metrics}
\subsubsection{Detecting human contraction waves}
\begin{figure}[t]
\centering
\includegraphics[width=8cm]{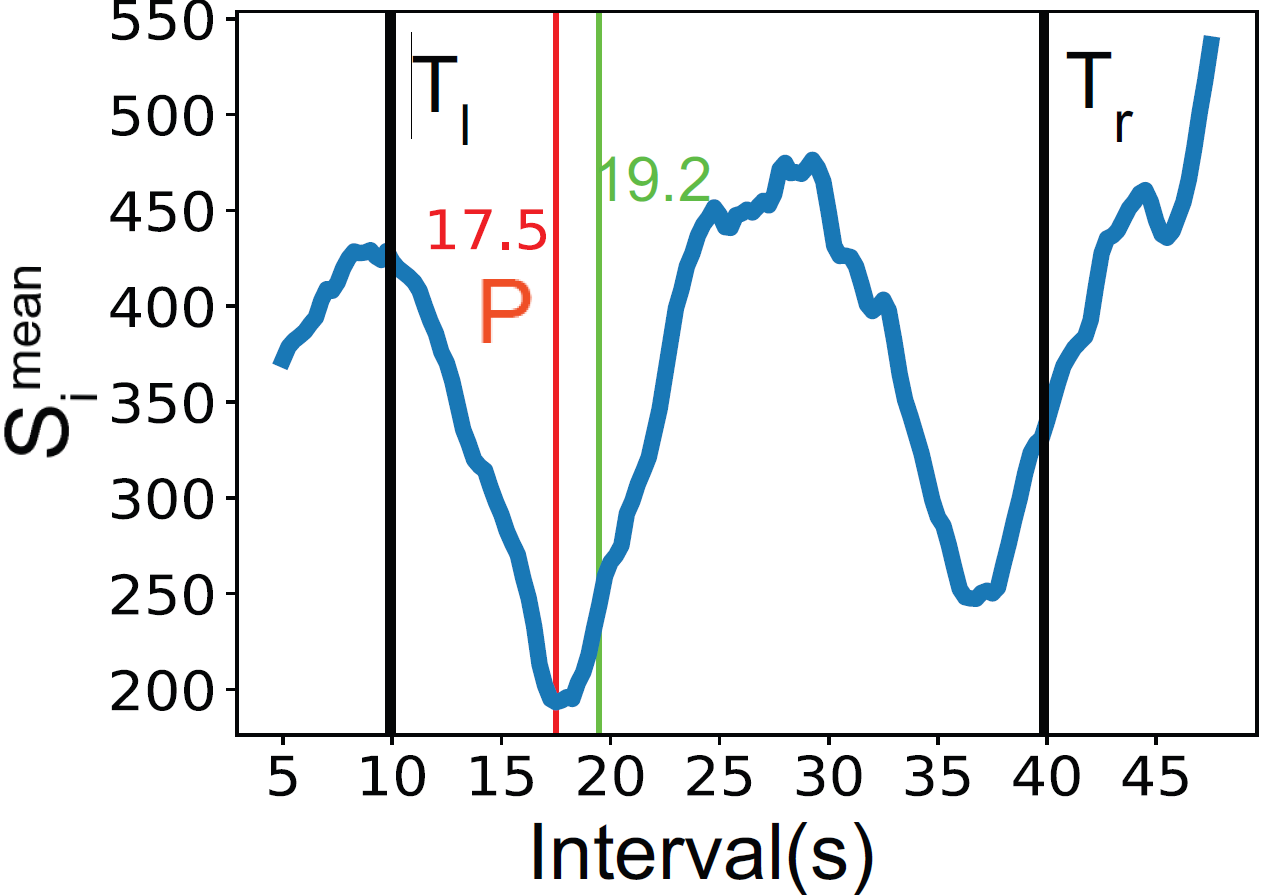}
\caption{An example of applying the periodical detector on an MCCE video sequence (case2 in the testing set). For each interval value $i$ from 5s to 50s, the periodical detector will generate a corresponding score $S^{mean}_{i}$. The $i$ (between predefined $T_l$ and $T_r$) with local minimum $S^{mean}_{i}$ will be identified as the detected period $P$. In this case, $P$ is 17.5$s$ (denoted as red line). The manually counted period (denoted as green line) is 19.2$s$}
\label{fig:LocalMinima}
\end{figure}
We process a large collection with more than 100,000 MCCE frames \textcolor{black}{from 30 subjects}. Using the CMD, we acquire 32,431 stable MCCE frames for training. For testing, we use \textcolor{black}{30 MCCE records from 11 subjects. The length of each testing record is longer than 50 seconds and independent from the training set}. Upon the completion of training, we evaluated the detection performance using accuracy, F1 score, and Area Under Curve (AUC) on the testing set. 
\subsubsection{Measuring frequency of human contraction waves}
\begin{figure*}[t]
\centering
\includegraphics[width=\textwidth]{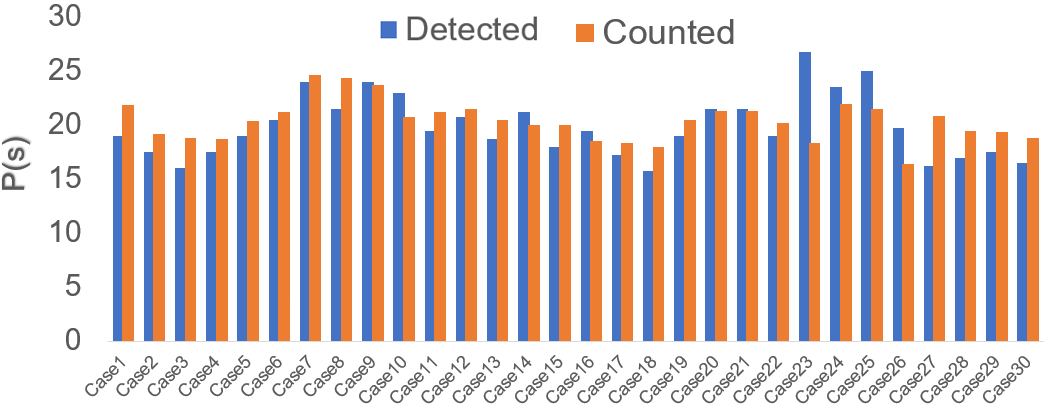}
\caption{Detected (denoted by blue color) and counted (denoted by orange) period of human contraction waves for the testing set of 30 MCCE video sequences. }
\label{fig:Period}
\end{figure*}
We apply the periodical detector to the testing set of 30 MCCE sequences. The detected periods, P, are compared with manually counted periods. \textcolor{black}{The ground truth is obtained from experienced readers.} For each MCCE sequence, the period of each contraction wave is counted starting from the first frame of the current wave to the first frame of the next wave. The average period of waves in a record is adopted as the manually counted period. We compare the automatically detected and manually counted periods by a defined error: $|$Detected period - Counted period$|$ / Counted period * 100$\%$.

\subsection{Results}
\subsubsection{Detecting human contraction waves}
\begin{table}
\centering
\caption{Performance of multiple variations of EfficientNet-B7+BiLSTM(x), where x represents length of  BiLSTM.
}
\label{table:tab1}
\begin{tabular}{cccc} 
\hline
\multicolumn{1}{l}{\diagbox{Methods}{Metrics}} & \multicolumn{1}{l}{Accuracy} & F1     & AUC     \\ 
\hline
EfficientNet-B7                                            & 0.8347                       & 0.7608 & 0.8964  \\
EfficientNet-B7+BiLSTM5                                      & 0.8570                       & 0.7840 & 0.9294  \\
EfficientNet-B7+BiLSTM10                                     & 0.8716                       & 0.7961 & 0.9335  \\
EfficientNet-B7+BiLSTM20                                     & 0.8882                       & 0.8192 & 0.9400  \\
\hline
\end{tabular}
\end{table}
\vspace{-6pt}
We train the EfficientNet-B7+BiLSTM model using the BiLSTM with lengths of 5, 10, and 20 for a comparative study. We apply the trained model with EfficientNet-B7 to detect the contraction waves. The results are shown in Table \ref{table:tab1}. We observed that longer lengths of BiLSTM leads to higher accuracy, F1, and AUC scores. This experiment demonstrates that features in the temporal domain are important for the detection of contraction waves in the MCCE sequences. 

\subsubsection{Measuring period of human contraction waves}
\begin{figure}[t]
\centering
\includegraphics[width=\columnwidth]{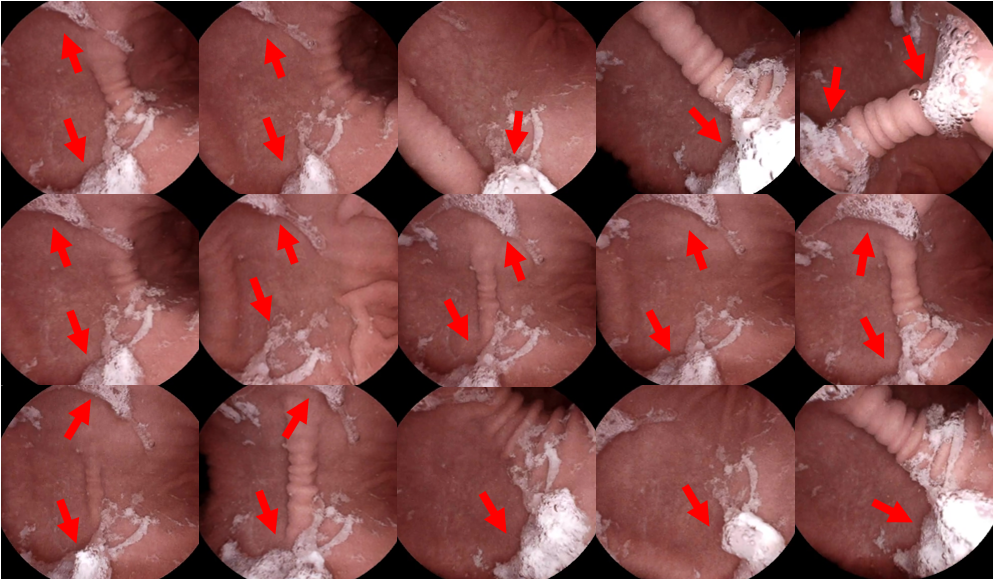}
\caption{Representative MCCE frames with the presence of mucus in case23 of the testing set. The mucus has shape features and motions different than the contraction waves, which may deteriorate the performance of the periodical detector. The red arrows highlight the clusters of mucus.}
\label{fig:castStudy}
\end{figure}
\vspace{-6pt}
To accurately extract the periods of contraction waves, we develop the periodical detector as described in Algorithm \ref{alg:cap2}. An example of using the periodical detector is shown in Figure \ref{fig:LocalMinima}. The periodical detector is capable of measuring the periods of contraction waves and providing results that are close to that of manually counting. In this case, we achieve an error of 8.85\%. The results of applying the periodical detector on the 30 MCCE records are shown in Figure \ref{fig:Period}. The mean, standard deviation (std), median, max, and min values of errors between the detected and counted period of contraction waves are shown in Table \ref{table:tab2}. Our periodical detector achieves a mean error of $10.05\%$ and a median error of $8.50\%$ on the 30 testing records. It is worth mentioning that in case23, a substantial amount of mucus is presented on the gastric, which might be the reason leading to a high error of $45.38\%$. We consider it as an outlier caused by a rare case of clustered mucus existence. Some representative MCCE frames with the presence of mucus are shown in Figure \ref{fig:castStudy}.
\begin{table}
\centering
\caption{The mean, standard deviation (std), median, max, and min values of the error between detected and counted period of human contraction waves for the testing set of 30 MCCE recordings.}

\label{table:tab2}
\begin{tabular}{lllll} 
\hline
\multicolumn{1}{c}{Mean} & \multicolumn{1}{c}{Std} & Median & \multicolumn{1}{c}{Max} & \multicolumn{1}{c}{Min}  \\ 
\hline
10.05\%                  & 8.27\%                  & 8.50\% & 45.38\%                & 0.93\%                   \\ 
\hline
                         &                         &        &                         &                         
\end{tabular}
\vspace{-10pt}
\end{table}

\section{CONCLUSIONS}
We have developed a CNN(EfficientNet-B7)+(Bi)LSTM model and a periodical detector for detecting and measuring periods of human gastric contraction waves captured by MCCE video sequences. Also, we have developed a CMD that is capable of processing MCCE video sequences. Our algorithms can work together during human gastric examination using MCCE, proving both qualitative (detection) and quantitative (period measuring) analysis of gastric motility. To the best of our knowledge, we are the first to propose computer vision-based solutions for studying human gastric motility.

In the future, we will improve the robustness of our algorithms, especially for cases where mucuses are present in the gastric. Also, we will 
extend our work by performing clinical experiments and recruiting subject groups with more diversity other than healthy. 
Moreover, we will extract more information, such as the frequency and amplitude, of gastric contraction waves to support clinical decision-making. 

\section{COMPLIANCE WITH ETHICAL STANDARDS}
This research was conducted retrospectively using MCCE data collected in the research department of Ankon Technologies and its collaboration hospitals. The MCCE data were de-identified. Written consent for reusing each data record was acquired from volunteers. IRB approval was not required as confirmed by the doctors.
\section{Acknowledgments}
This research was sponsored by AnX Robotica. This work was done during Xueshen Li's internship at AnX Robotica. Xiao Yang and David Duan are employed by AnX Robotica. The views expressed are those of the authors and do not necessarily reflect the official opinions or views of AnX Robotica. The algorithms described in this work are not official products of AnX Robotica. The authors declare no conflicts of interest. The correspondence author: xiao.yang@anxrobotics.com.
\label{sec:acknowledgments}
\bibliographystyle{IEEEbib}
\bibliography{strings,refs}

\end{document}